\documentclass[fleqn,usenatbib]{mnras}

\usepackage[T1]{fontenc}
\usepackage{ae,aecompl}

\usepackage{graphicx}	
\usepackage{amsmath}	
\usepackage{amssymb}	

\title[Dynamical evolution of fractals]{Dynamical evolution of fractal structures in star-forming regions}

\author[E. C. Daffern-Powell \& R. J. Parker]{
Emma C. Daffern-Powell,\thanks{E-mail: ecdaffern1@sheffield.ac.uk}
Richard J. Parker\thanks{Royal Society Dorothy Hodgkin fellow}
\\
Department of Physics and Astronomy, The University of Sheffield, Hicks Building, Hounsfield Road, Sheffield S3 7RH, UK
}

\date{Accepted XXX. Received YYY; in original form ZZZ}

\pubyear{2019}

\begin{document}
\label{firstpage}
\pagerange{\pageref{firstpage}--\pageref{lastpage}}
\maketitle

\begin{abstract}
  The $\mathcal{Q}$-parameter is used extensively to quantify the spatial distributions of stars and gas in star-forming regions as well as older clusters
  and associations. It quantifies the amount of structure using the ratio of the average length of a minimum spanning tree, $\bar{m}$, to the average length
  within the complete graph, $\bar{s}$. The interpretation of the $\mathcal{Q}$-parameter often relies on comparing  observed values of $\mathcal{Q}$, $\bar{m}$,
  and $ \bar{s}$ to idealised synthetic geometries, where there is little or no match between the observed star-forming regions and the synthetic regions. 
We measure $\mathcal{Q}$, $\bar{m}$, and $ \bar{s}$ over 10 Myr in $N$-body simulations which are compared to IC 348, NGC 1333, and the ONC. For each star-forming region we set up simulations that approximate their initial conditions for a combination of different virial rations and fractal dimensions. We find that dynamical evolution of idealised fractal geometries can account for the observed $\mathcal{Q}$, $\bar{m}$, and $ \bar{s}$ values in nearby star-forming regions. In general, an initially fractal star-forming region will tend to evolve to become more smooth and centrally concentrated. However, we show that initial conditions, as well as where the edge of the region is defined, can cause significant differences in the path that a star-forming region takes across the $\bar{m}-\bar{s}$ plot as it evolves.
We caution that the observed $\mathcal{Q}$-parameter should not be directly compared to idealised geometries. Instead, it should be used to determine the degree to which a star-forming region is either  spatially substructured or smooth and centrally concentrated. 

\end{abstract}




\section{Introduction} 

The majority of stars form in filamentary structures within giant molecular clouds, where the stellar density exceeds that of the galactic field \citep[e.g.][]{2003LadaLada, 2012GielesEtAl, 2014AndreEtAl}.
Even the least dense of these star-forming regions have stellar densities of a few stars pc$^{-3}$ (Taurus has $\approx5$ stars pc$^{-3}$) compared to 0.1 pc$^{-3}$ for the field \citep{2003KorchaginEtAl, 2012KingEtAl}. 
Some star-forming regions can have densities as high as $\gtrsim1000$ stars pc$^{-3}$, with the Orion Nebula Cluster having a stellar density of $\approx5000$ pc$^{-3}$ \citep{2012KingEtAl}.

These stellar densities can have significant effects on star and planet formation. 
Star-forming regions have the potential to perturb and destroy planetary \citep{2012ParkerQuanz, 2016KouwenhovenEtAl} and multiple stellar systems \citep{1995Kroupa1491, 1995Kroupa1507, 2011ParkerEtAl, 2012MarksKroupa}, to truncate and destroy protoplanetary disks \citep{2011WilliamsCieza, 2019NicholsonEtAl},
and may have even affected the early Solar System \citep{2010Adams}. For example, isotope enrichment from a nearby supernova may have decreased planetary water abundances \citep{2019LichtenbergEtAl}, and dynamical interactions with a passing star may have shaped the outer Solar System \citep{2018PfalznerEtAl}.
It is therefore important to be able to quantify the properties of star-forming regions, in order to compare simulations to observations and to better understand the effects that these environments can have.

There is evidence that star-forming regions tend to be substructured, both initially and for a time during their evolution \citep{2004CartwrightWhitworth, 2009SanchezAlfaro, 2010AndreEtAl, 2014AndreEtAl, 2014KuhnEtAl, 2015JaehnigEtAl, 2019ArzoumanianEtAl, 2020BalloneEtAl}.
Areas of substructure tend to have relatively high stellar densities compared to the region as a whole, and can therefore be more detrimental to star and planet formation \citep[e.g.][]{2012ParkerQuanz}.

There are several methods that can be used to measure and quantify substructure. 
These methods can often be used to identify the star-forming regions themselves, as well as any substructure at different scales within them \citep{2011Schmeja}.
The most basic method is a stellar density map, where a region is split into bins and the number of stars in each bin is compared in order to identify areas that are significantly over-dense \citep{2011Schmeja}.

More complex methods include the nearest neighbour density, which estimates the local density around each star \citep{2011Schmeja, 2009GutermuthEtAl, 2019BucknerEtAl};
the angular dispersion parameter, which divides a region into segments and compares their stellar densities \citep{2014DaRioEtAl,2015JaehnigEtAl};
the two-point correlation function, which identifies pairs of stars that are closer than average \citep{1993GomezEtAl, 1995Larson, 2014GouliermisEtAl};
and the minimum spanning tree, which is commonly used to identify substructure \citep{2004CartwrightWhitworth, 2009GutermuthEtAl, 2011KirkMyers, 2011Schmeja}.

In a minimum spanning tree, all of the stars in a region are connected such that the total length of all the edges (i.e. connections) is minimised and there are no closed loops. 
Sub-clusters and areas of substructure can then be identified by removing edges which are longer than a chosen length \citep{2009GutermuthEtAl, 2011KirkMyers, 2011Schmeja}. 
And the mean edge length, $\overline{m}$, can be used to quantify how substructured or centrally concentrated a region is \citep{2004CartwrightWhitworth}. 

However, $\overline{m}$ alone is unable to distinguish between substructured and smooth centrally concentrated regions \citep{2004GoodwinWhitworth}.
To overcome this, the $\mathcal{Q}$-parameter was introduced by \citet{2004CartwrightWhitworth}, and further developed by \citet{2009Cartwright}, \citet{2011LomaxEtAl}, and \citet{2017JaffaEtAl}.
The $\mathcal{Q}$-parameter is calculated using equation~\ref{eq:Q}:
\begin{equation}
    Q = \frac{\overline{m}}{\overline{s}}
    \label{eq:Q}
\end{equation}
Here, $\overline{s}$ is the mean edge length of the star-forming region's complete graph, where edges are drawn from every star to every other star \citep{2009Cartwright}.

As $\overline{m}$ and $\overline{s}$ are two different measures, they scale differently. 
This enables $\mathcal{Q}$ to distinguish between substructured and centrally concentrated regions, as it combines $\overline{m}$ and $\overline{s}$. 
The $\mathcal{Q}$-parameter therefore both gives a measure of the amount of substructure and distinguishes substructured regions from those that are smooth and centrally concentrated, in a single dimensionless number. 
Regions with $Q \gtrsim 0.8$ have a smooth radial density profile that is more centrally concentrated with higher $\mathcal{Q}$, while regions with $Q \lesssim 0.8$ have a larger amount of substructure with decreasing $\mathcal{Q}$. 
The $\mathcal{Q}$-parameter has been used extensively for investigating substructure in both simulated and observed regions \citep[e.g.][]{2006SchmejaKlessen, 2009BastianEtAl, 2009SanchezAlfaro, 2013DelgadoEtAl, 2014ParkerEtAl, 2015ParkerDale}.

\citet{2009Cartwright} showed that using a plot of $\overline{m}$ vs. $\overline{s}$ provides more information than $\mathcal{Q}$ alone, as it contains more information than a single number, and is therefore more sensitive to distinguishing between different properties. The interpretation of the $\overline{m}$ vs. $\overline{s}$ plot relies on comparing the observed or simulated values to sets of idealised geometries, which are usually either box fractals or centrally concentrated spheres.

\begin{figure*}
    \label{fig:msplotgap}
    \centering
    \includegraphics[width=\textwidth]{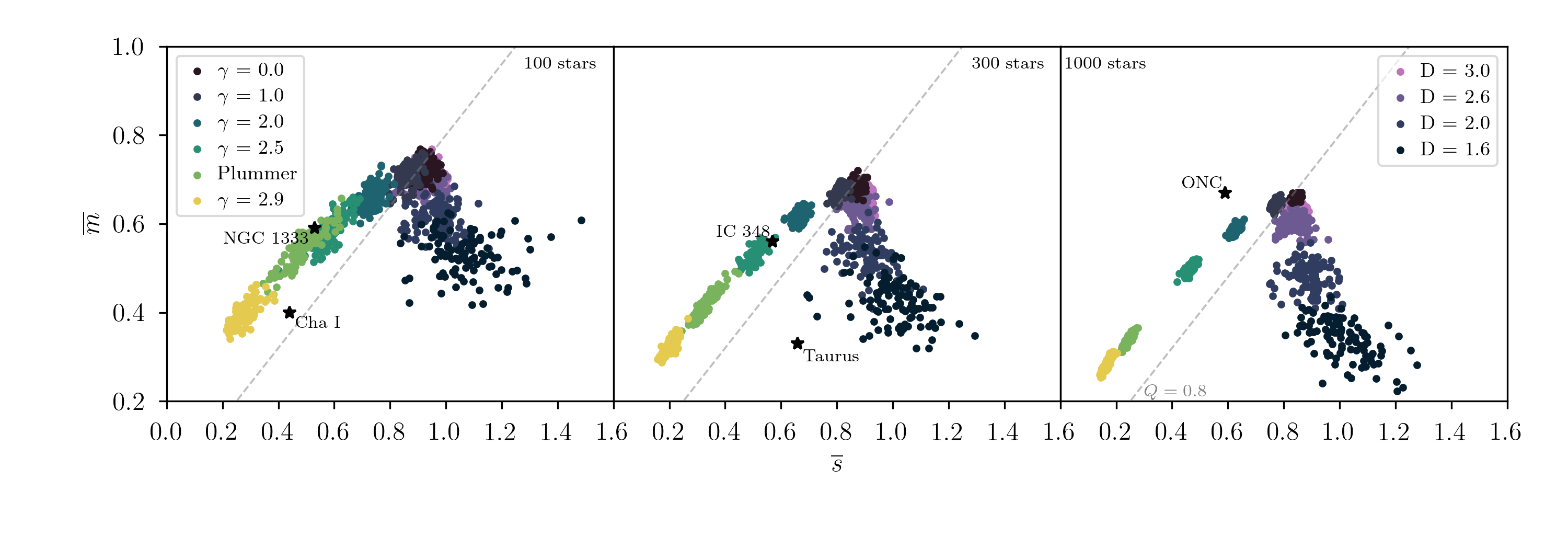}
    \caption{A comparison between the areas of the $\overline{m}-\overline{s}$ plot that are 	occupied by observed regions vs. idealised geometries.
    For the observed regions, values of $\overline{m}$ and $\overline{s}$ for IC~348 \citep{2017ParkerAlvesDeOliveira}, NGC~1333 \citep{2017ParkerAlvesDeOliveira}, the ONC \citep[data from][]{1997Hillenbrand}, Cha I \citep[data from][]{2007Luhman}, and Taurus \citep[data from][]{2010LuhmanEtAl}, are marked by black stars.
    For the idealised geometries, box fractal regions and those with a smooth radial density profile are shown as coloured points. 100 different realisations of each type of idealised region are shown. These regions were created with either 100, 300, or 1000 stars, as shown in the 3 panels. The observed regions are therefore shown on the panel that approximately corresponds to their observed number of stars.
    Fractal dimensions of $D = 3.0, 2.6, 2.0$, and $1.6$ are shown, with the regions becoming more substructured with lower values of $D$. 
    For the centrally concentracted regions, a Plummer Sphere \citep{1911Plummer} and regions with radial density profile exponents of $\gamma = 0.0, 1.0, 2.0, 2.5$ and $2.9$ are shown, with the regions becoming more centrally concentrated with lower values of $\gamma$.
    A gap can be seen between the box fractals and radial density profiles which is occupied by observed regions, for example Cha I and Taurus, meaning that these idealised geometries do not describe the observed substructure in these regions.}
    \label{fig:my_label}
\end{figure*}

However, values of $\overline{m}$ and $\overline{s}$ for observed star-forming regions often do not occupy the same areas of the $\overline{m}$-$\overline{s}$ plot as idealised box fractals and smooth centrally concentrated regions \citep{2018LomaxBatesWhitworth}. This is shown in Figure~\ref{fig:msplotgap}, where a clear gap can be seen in between the box fractal regions and those with a smooth radial density profile. However, this area is still populated by observed regions, e.g Cha I and Taurus.

It has been suggested that this is a shortcoming of the methods themselves, as it can be seen to imply that $\overline{m}$ and $\overline{s}$ are not able to characterise observed regions and/or that these idealised geometries are not reasonable approximations for star-forming regions \citep{2018LomaxBatesWhitworth}. 

In this paper, we use the $\overline{m}-\overline{s}$ plot to test whether observed star-forming regions are consistent with having evolved from fractal geometries, or whether there is a problem with the ability of the $\mathcal{Q}$-parameter to quantify substructure. We outline our methods in Section~2, we show our results in Section~3, we provide a discussion in Section~4 and we conclude in Section~5.

\section{Methods}
We simulate regions that approximate IC~348, NGC~1333, and the ONC using the \texttt{kira} N-body integrator \citep[e.g.][]{1999PortegiesZwart, 2001PortegiesZwart}. 

Table~\ref{tab:initials} shows the initial conditions for each set of simulations.
With regards to the choice of initial conditions, $\mathcal{Q}$ is dependant on the number of points in the distribution \citep[see][his Fig. A2]{2018Parker}, and is also significantly affected by the inclusion of foreground/ background stars. It is therefore important to only include stars with a high membership probability in the analysis of observational data, and to run simulations using an equal number of stars to this observational sample so as to allow for a direct comparison. 
For example, for the ONC, there are 929 stars with a $>90\%$ membership probability \citep{1997Hillenbrand, 1998HillenbrandHartmann, 2011ReggianiEtAl}, we therefore run simulations with, $N_{\star} = 929$. 
Similarly, for IC 348 and NGC 1333, there are respectively 459 and 162 stars with  membership  confirmed by \citet{2016LuhmanEtAl}.
This inevitably means that some genuine members may be excluded from the analysis, with fainter, lower mass members being disproportionately affected. However, it is not expected that $\mathcal{Q}$ varies as a function of stellar mass such that it would be affected by this \citep{2014ParkerEtAl}. It is therefore a fairer comparison to the data to use the lower values of $N_{\star}$ adopted here, even for clusters which likely have uncatalogued members, as may be the case for the ONC \citep[e.g.][]{2016ForbrichEtAl}.

We inferr initial radii, $R$, from comparing the amount of mass segregation in simulations to the observed levels of mass segregation in each of the ONC \citep{2010AllisonEtAl, 2011AllisonGoodwin}, IC~348, and NGC~1333 \citep{2017ParkerAlvesDeOliveira}. \citet{2014ParkerEtAl} show that the level of mass segregation in a star-forming region is a proxy for the amount of dynamical evolution that has taken place, which in turn  places constraints on the initial density (and therefore radius). If future observations add significant numbers of extra stars (i.e.\,\,a factor of two more) to the regions' censuses, then our analyses would need to be repeated with new simulations better tailored to the observed numbers of stars. 

These radii are combined with different initial fractal dimensions, $D$, and virial ratios, $\alpha$. The stellar masses are sampled from a Maschberger IMF \citep{2013Maschberger}, with minimum and maximum masses of 0.1 M$_\odot$ and 50 M$_\odot$ respectively.

For each set of initial conditions in Table~\ref{tab:initials}, ten realisations are simulated using different random number seeds. Each simulation is evolved for 10 Myr. We do not include stellar evolution or primordial binaries in the simulations.

\begin{table}
	\centering
	\caption{Ages and initial conditions used for each set of simulations. 
	Columns 2 and 3 contain the current observed ages and number of stars for IC~348 \citep{2016LuhmanEtAl}, NGC~1333 \citep{2016LuhmanEtAl}, and the ONC \citep{1997Hillenbrand, 1998HillenbrandHartmann, 2011ReggianiEtAl}.
	Column 4 gives the initial radii, $R$, as inferred from simulations \citep{2010AllisonEtAl, 2011AllisonGoodwin}.
    Columns 5 and 6 show the combinations of fractal dimension, $D$, and virial ratio, $\alpha$, used here.}
	\label{tab:initials}
	\begin{tabular}{l c c c c c}
		\hline
		Name     & Age/Myr      & $N_{\star}$ & $R$/pc & $D$  & $\alpha$ \\
		\hline 
		IC~348   & $\sim$3      & 459         & 1.5    &  1.6 & 0.3 \\
		         &              &             &        &  1.6 & 1.5 \\
		         &              &             &        &  2.0 & 0.3 \\
		         &              &             &        &  2.0 & 1.5 \\
		         &              &             &        &  3.0 & 0.3 \\
		NGC~1333 & $\sim$1      & 162         & 0.5    &  1.6 & 0.3 \\
		         &              &             &        &  1.6 & 1.5 \\
		         &              &             &        &  2.0 & 0.3 \\
		         &              &             &        &  3.0 & 0.3 \\
		ONC      & $\approx$1-4 & 929         & 1      &  1.6 & 0.3 \\
		         &              &             &        &  1.6 & 1.5 \\
		         &              &             &        &  2.0 & 0.3 \\
		         &              &             &        &  3.0 & 0.3 \\
		\hline
	\end{tabular}
\end{table}

\subsection{Simulations}
\label{Method_Simulations}

\subsubsection{Spatial Substructure}
The initial substructure is set up using a box fractal distribution. The box fractal method is a commonly used and convenient way of producing substructure, partly because the amount of substructure is defined by one number: the fractal dimension, $D$.

The spatial distribution is set up using the method outlined in \citet{2004GoodwinWhitworth}:
\renewcommand{\labelenumi}{\arabic{enumi})}
\begin{enumerate}
    \item A cube with sides of length $N_{\rm div} = 2$ is defined, within which the region is to be generated. The first `parent' star is placed at its centre.
    \item This cube is divided into $N_{\rm div}^3$ sub-cubes, and a `child' star is placed at the centre of each sub-cube. So, in this case, there are 8 sub-cubes.
    \item The probability that a child now becomes a parent itself is $N_{\rm div}^{(D-3)}$.
    \item Children who do not become parents are removed, as well as all of their parent stars.
    \item Children who do become parents have a small amount of noise added to their positions, to prevent a gridded appearance. 
    \item Each child's sub-cube is then divided into $N_{\rm div}^3$ itself, as the process is repeated until there is a generation with significantly more stars than needed.
    \item Any remaining parents are removed, so that only the last generation is left.
    \item The region is pruned so that the stars sit within the boundary of a sphere, rather than a  cube. 
    \item If there are more stars remaining than the specified $N_{\star}$, stars are removed at random until $N_{\star}$ is reached. This maintains the chosen fractal dimension as closely as possible.
\end{enumerate}

The mean number of children that become parents is $N_{\rm div}^{D}$. So, when $N_{\rm div} = 2$, fractal dimensions of $D =$ 1.6, 2.0, 2.6, and 3.0 correspond to the mean number of new parents at each stage being close to an integer. This is preferred because it produces the chosen fractal dimension more accurately. Meanwhile, a lower fractal dimension leads to fewer children becoming parents, and therefore more substructure. So, here, $D = 1.6$ is the maximum amount of substructure possible, and $D = 3.0$ produces a uniform, non-substructured, distribution because all of the children become parents. We adopt values of $D =$ 1.6, 2.0, and 3.0 for the initial conditions in our simulations.

\subsubsection{Stellar Velocity}
The stellar velocities are substructured,  which means that stars that are closer together have more similar velocities than those that are further apart. This is also done according to the method in \citet{2004GoodwinWhitworth}: 
\begin{enumerate}
    \item The first parent star has its velocity drawn from a Gaussian with mean zero. 
    \item Every star after that has the velocity of its parent plus an additional random velocity component. This additional component is drawn from the same Gaussian and multiplied by $(\frac{1}{N_{\rm div}})^{g}$, where $g$ is the generation that the star was produced through the box fractal method. This results in the additional components being smaller on average with each successive generation of stars created.
    \item The velocities are scaled so that the region has the required virial ratio, $\alpha = T / |\Omega|$, where $T$ is the total kinetic energy of the region, and $\Omega$ is the total potential energy.
\end{enumerate}

Here, virial ratios of $\alpha =$ 0.3, and 1.5 are used. Where regions with $\alpha = 0.3$ are initially subvirial and in cool-collapse, and regions with $\alpha = 1.5$ are initially supervirial and expanding.

\subsection{Determining Q}
\label{sec:Method_Qpar}

For each simulation, $\overline{m}$ and $\overline{s}$ are calculated every 0.01 Myr. This was done for different normalisation methods and cut-off radii, as described in the following sub-sections.

For all calculation methods, $\overline{m}$ and $\overline{s}$ were calcuated in 2D, to mimic a projection on the sky, allowing a more direct comparison to observations.

\subsubsection{Cut-Off Boundary}
Three different membership criteria were used to determine which stars should be included in the calculation of $\overline{m}$ and $\overline{s}$. The first criterion is simply to include all of the stars in the simulation in the analysis. Two cut-off boundaries are also used, beyond which stars are excluded from the analysis as they may not be observationally associated with the region based on their distance from its centre. These cut-off radii were chosen to be 5pc and 3pc.

\subsubsection{Normalisation}
Both $\overline{m}$ and $\overline{s}$ must be normalised with respect to the region's size.
For $\overline{s}$, the mean edge length of the complete graph is normalised to the region's radius \citep{2004CartwrightWhitworth}.
For $\overline{m}$, the mean edge length of the minimum spanning tree is normalised with respect to the region's area \citep{2004CartwrightWhitworth}, by dividing by a factor of:
\begin{equation}
    \frac{\sqrt{NA}}{N - 1},
\end{equation}
where $N$ is the number of points, and $A$ is the region's projected area on the sky.

The characteristic area of a region can either be taken to be that of a circle, or a convex hull - an enclosure drawn around the outermost stars, so that all of the stars are enclosed and the total length of the edges of the enclosure is minimised. Here, $\overline{m}$ and $\overline{s}$ were calculated using both circular and convex hull normalisations for comparison, as the normalisation can have a significant effect on the results \citep{2018Parker}.

For the standard circular normalisation, the area is that of a circle, with radius drawn from the region's centre to its outermost star \citep{2004CartwrightWhitworth}. For the convex hull normalisation we use the method introduced by \citet{2006SchmejaKlessen}, where $\overline{m}$ is normalised to the area of the convex hull, and $\overline{s}$ is normalised to the radius of a circle that has the same area as that convex hull.

\section{Results} 
Our main results are shown in Figures 2-10. 
These figures show the evolutionary tracks of our simulations across the $\overline{m}-\overline{s}$ plot, along with the current observed values of $\overline{m}$ and $\overline{s}$ for their corresponding regions shown as a yellow star.
We discuss these figures in turn in the following subsections.

\subsection{IC~348} 
\subsubsection{$D = 2.0$, $\alpha = 0.3$}
\begin{figure*}
	\includegraphics[width=\textwidth]{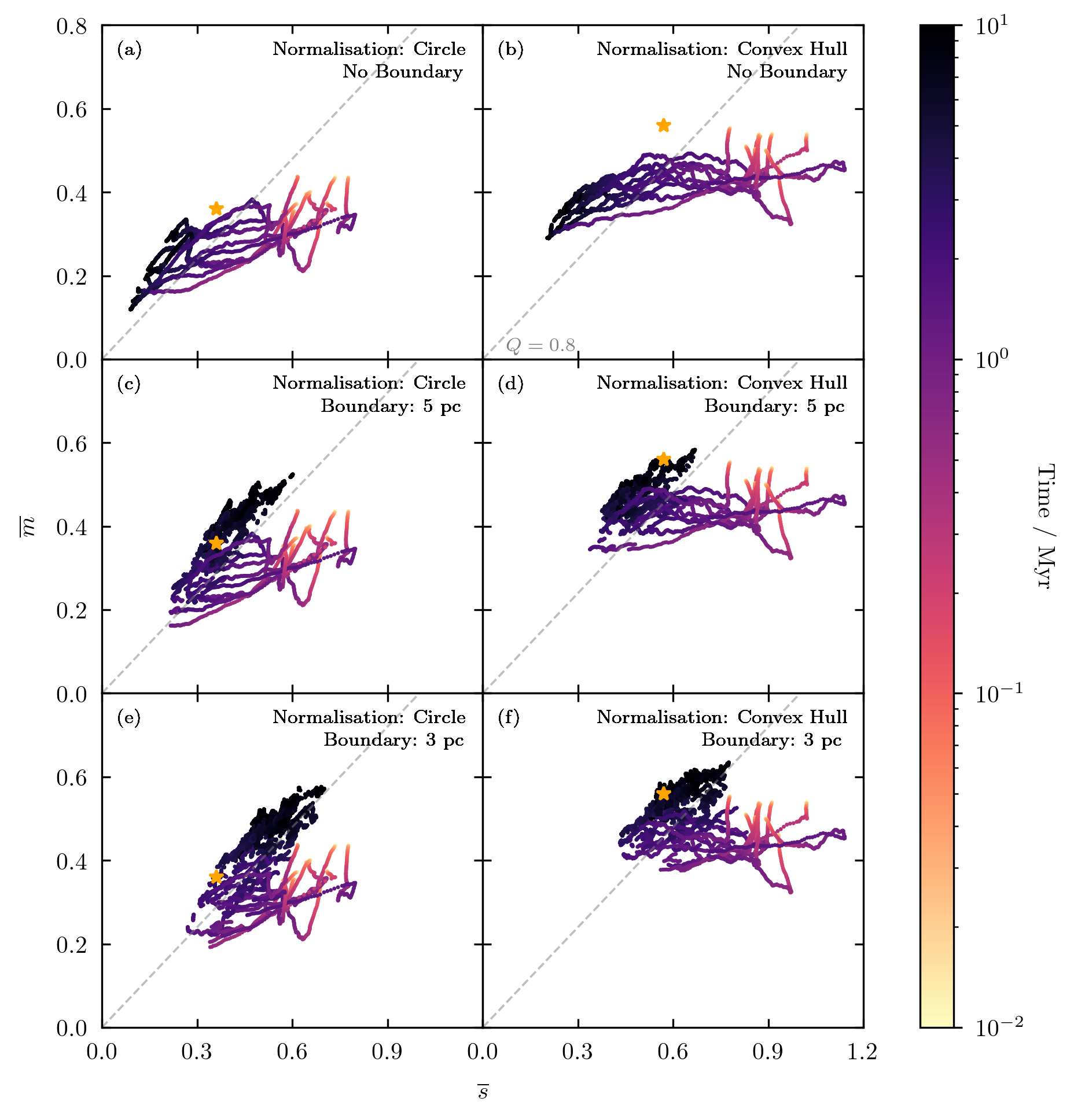}
    \caption{Evolution of $\overline{m}$ and $\overline{s}$ over 10 Myr for an IC~348-like star-forming region, with initial $D = 2.0$ and $\alpha = 0.3$.
    Ten realisations of the region are shown for 6 combinations of analysis methods.
    The observed values of $\overline{m}$ and $\overline{s}$ for IC~348 (age $\approx3-6$ Myr) are shown as a yellow star, and a grey dashed line shows the $Q = 0.8$ boundary between substructured and centrally concentrated distributions.
    Each realisation crosses the $\overline{m}-\overline{s}$ plot gap as they dynamically evolve during the first $\sim1$ Myr.}
    \label{fig:ms_IC348_D2_all}
\end{figure*}

Figure~\ref{fig:ms_IC348_D2_all} shows the evolution of each realisation of the IC~348-like $\alpha = 0.3$ $D = 2.0$ simulations, for all three region membership criteria and both normalisation methods.

Panels (a) and (b) show that, when all stars are used in the analysis, the overall evolution of each realisation is a rapid ($\sim0.1$ Myr) initial drop in $\overline{m}$, followed by a movment towards lower values of $\overline{m}$ and often $\overline{s}$. This movement crosses the gap, shown in Figure~\ref{fig:msplotgap}, demonstrating that these regions can populate the $\overline{m}-\overline{s}$ plot gap as they dynamically evolve.

\begin{figure}
    \centering
    \includegraphics[width=0.7\columnwidth]{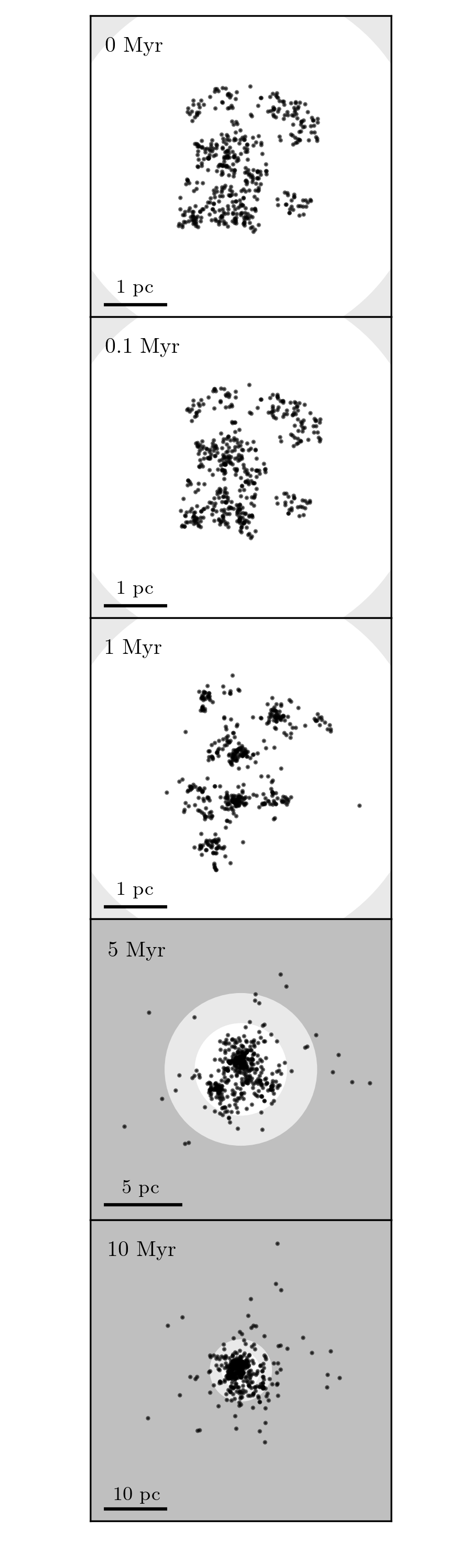}
    \caption{x-y stellar positions, at 5 different times, showing the dynamical evolution of one realisation of the IC~348-like star-forming region with initial $D = 2.0$ and $\alpha = 0.3$.
    Stars are shown as slightly transparent black dots, so that denser areas are more opaque.
    The area beyond the 5 pc cut-off is shaded dark grey, and the area in between the 5 pc and 3 pc cut-offs is shaded a lighter grey.
    Initially, between $\sim0-1$ Myr, the individual areas of substructure collapse. This is followed by the region as a whole dynamically interacting, causing it to transition to a centrally concentrated geometry and eject stars.}
    \label{fig:xy_overtime}
\end{figure}

A comparison to Figure~\ref{fig:msplotgap} shows that the initial drop corresponds to the regions becoming more substructured. This is counter-intuitive, as dynamical interactions erase substructure. However, this phase corresponds to the `clumps' of substructure within the regions collapsing on local scales, before the region as a whole has begun to collapse significantly. The reason why this is seen as an increase in substructure by the $\mathcal{Q}$-parameter is best understood visually. Figure~\ref{fig:xy_overtime} shows the spatial distribution of one realisation as it evolves over the 10 Myr simulation. Between $\sim 0 - 1$ Myr the region begins to collapse on local scales, and the clumps appear more pronounced as they become smaller and more centrally concentrated - it is this behaviour that is seen as an increase in the degree of substructure.

The subsequent movement towards lower $\overline{s}$ then corresponds to the region as a whole collapsing. Dynamical interactions wipe out substructure on global scales as the clumps are destroyed and the region as a whole becomes smoother and more centrally concentrated, as seen from the 1 - 10 Myr panels of Figure~\ref{fig:xy_overtime}. During this time, each realisation crosses the $Q \simeq 0.8$ boundary, so that all ten end in the area of the $\overline{m}-\overline{s}$ plot that corresponds to a centrally concentrated region with a smooth radial density profile.

When a cut-off boundary is imposed this behaviour changes, as shown in panels (c)-(f) of Figure~\ref{fig:ms_IC348_D2_all}. For both a 5 and 3 pc cut-off, each realisation still has an initial drop in $\overline{m}$ up to $\sim0.1$ Myr, followed by a migration towards lower $\overline{s}$. However, this migration stops at $\sim1$ Myr, and the final stage of the regions' evolution is a diagonal increase in $\overline{m}$ and $\overline{s}$. This happens with approximately constant $\mathcal{Q}$, just above the $Q =  0.8$ line, and means that the simulations cross the area of the $\overline{m}-\overline{s}$ plot where IC~348 is observed to be. This final stage begins once some of the stars reach the cut-off boundary, as the region expands. This expansion causes $\overline{m}$ and $\overline{s}$ to increase as the stars within the cut-off get farther apart, and fewer stars are left within it.

\begin{figure}
    \centering
    \includegraphics[width=0.9\columnwidth]{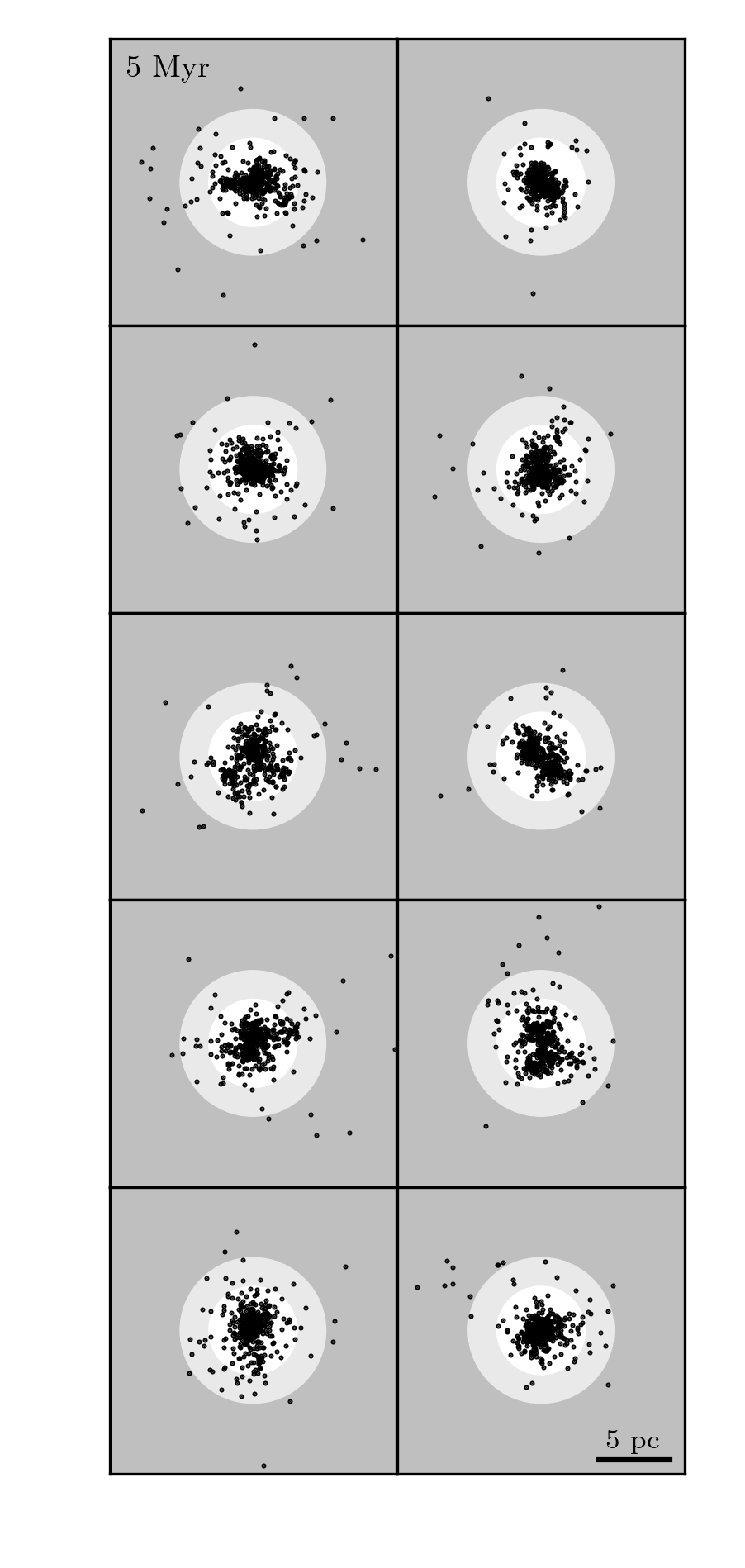}
    \caption{x-y stellar positions, at 5 Myr, for ten realisation of IC~348-like regions with initial $D = 2.0$ and $\alpha = 0.3$. Stars are shown as slightly transparent black dots, so that denser areas are more opaque.
    The area beyond the 5 pc cut-off is shaded dark grey, and the area in between the 5 pc and 3 pc cut-offs is shaded a lighter grey.
    Initially, between $\sim0-1$ Myrs, the individual areas of substructure collapse. This is followed by the region as a whole dynamically interacting, causing it to transition to a centrally concentrated geometry and eject stars.}
    \label{fig:xy_many}
\end{figure}

Figure~\ref{fig:xy_many} shows that, after 5 Myr, a 3 pc cut-off begins to exclude some stars which could reasonably be identified as belonging to the region based on their x-y positions, where a 5 pc cut-off does not. However, comparison to Figure~\ref{fig:ms_IC348_D2_all} shows that this does not have a significant effect. For the remaining figures, results are therefore shown with only a 5pc cut-off.

When the standard circular normalisation is used, compared to a convex hull, each realisation has lower values of $\overline{m}$ and $\overline{s}$. This is because a convex hull will always have a smaller area than the corresponding circle. The area and radius used to normalise $\overline{m}$ and $\overline{s}$ will therefore be higher for the circular normalisation method compared to convex hull normalisation \citep[see][]{2018Parker}.

\subsubsection{Effect of initial fractal dimension} 
\begin{figure*}
	\includegraphics[width=\textwidth]{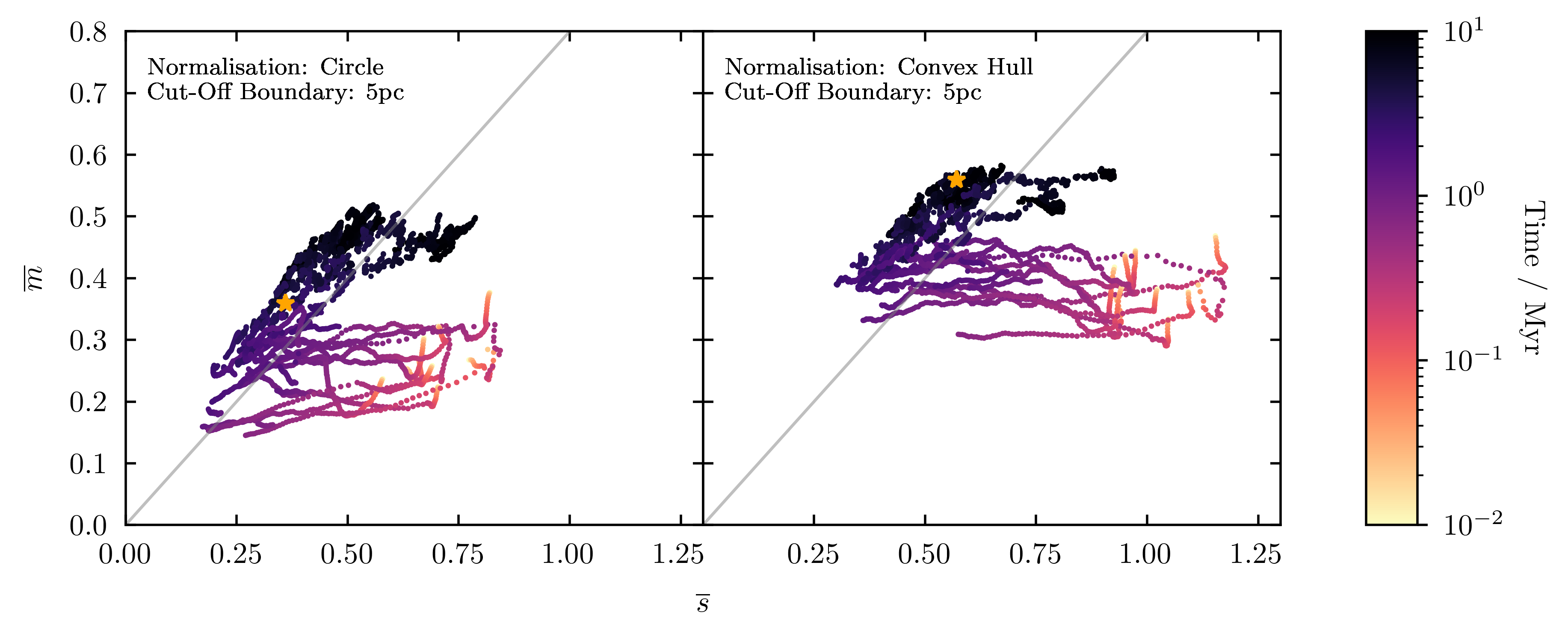}
	\caption{Evolution of $\overline{m}$ and $\overline{s}$ over 10 Myr for an IC~348-like region, with initial $D = 1.6$ and $\alpha = 0.3$.
    Ten realisations of the region are shown for both a circular and convex hull normalisations.
    The current values of $\overline{m}$ and $\overline{s}$ for IC~348 (age $\approx3-6$ Myr) are shown as a yellow star.
    The evolution is similar to the corresponding simulations with initial $D = 2.0$.}
    \label{fig:ms_IC348_D16_two}
\end{figure*}

\begin{figure*}
	\includegraphics[width=\textwidth]{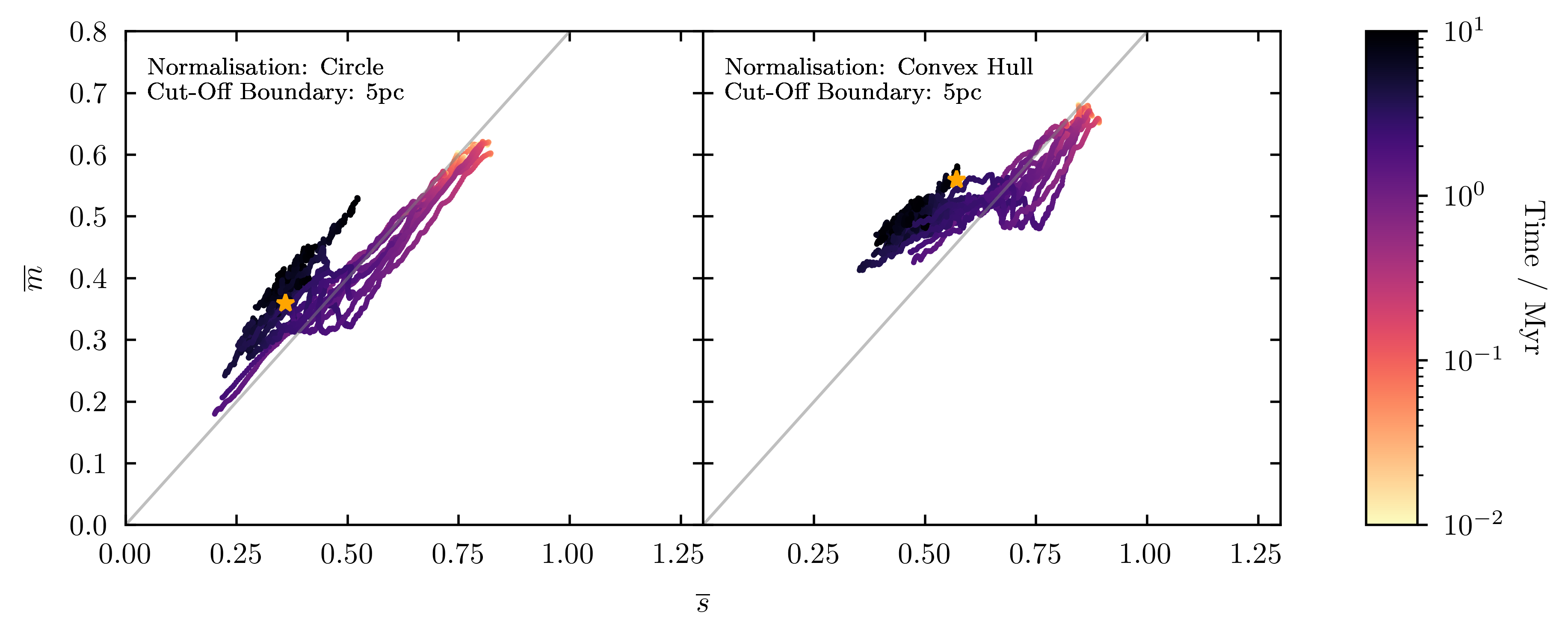}
	\caption{Evolution of $\overline{m}$ and $\overline{s}$ over 10 Myr for an IC~348-like region, with initial $D = 3.0$ and $\alpha = 0.3$.
    Ten realisations of the region are shown for both circular and convex hull normalisations.
    The current values of $\overline{m}$ and $\overline{s}$ for IC~348 (age $\approx3-6$ Myr) are shown as a yellow star.}
    \label{fig:ms_IC348_D3_two}
\end{figure*}

Figures~\ref{fig:ms_IC348_D16_two} and~\ref{fig:ms_IC348_D3_two} how the evolutions of the IC~348-like simulations differ with fractal dimension. 

Regions with a fractal dimension of $D = 1.6$ have the maximum amount of initial substructure, and therefore begin more clumpy than those with $D = 2.0$ discussed in the previous subsection. A comparison of Figures~\ref{fig:ms_IC348_D2_all} and~\ref{fig:ms_IC348_D16_two} shows that the evolution of the IC~348-like simulations is similar with initial $D = 1.6$ and $D = 2.0$. The $D = 1.6$ realisations begin lower on the $\overline{m}-\overline{s}$ plot, due to their larger amount of substructure, but migrate to the same area of the plot before making the upward turn at $\sim1$ Myr.

Regions with a fractal dimension of $D = 3.0$ are initially smooth and non-substructured.
Figure~\ref{fig:ms_IC348_D3_two} shows that these simulations also evolve to the same area of the plot as regions with initial $D = 2.0$ and 1.6, before the upwards turn at $\sim1$ Myr. 
In this case, this leads to each realisation initially migrating downward along the $Q=0.8$ line, corresponding to the regions remaining smooth, but becoming more centrally concentrated up to $\sim1$ Myr as they collapse. They then become less centrally concentrated as they expand past the 5 pc cut-off boundary for the remainder of the simulation. 

This means that this set of non-substructured regions do not cross the $\overline{m}-\overline{s}$ plot gap during their evolution, and would therefore be unable to explain regions which are observed to lie in the $\overline{m}-\overline{s}$ plot gap. However, since IC~348 is currently observed to lie in the smooth and centrally concentrated area of the $\overline{m}-\overline{s}$ plot, it is possible that IC~348 had an initially smooth distribution. 

\subsubsection{Effect of initial virial ratio} 
\begin{figure*}
	\includegraphics[width=\textwidth]{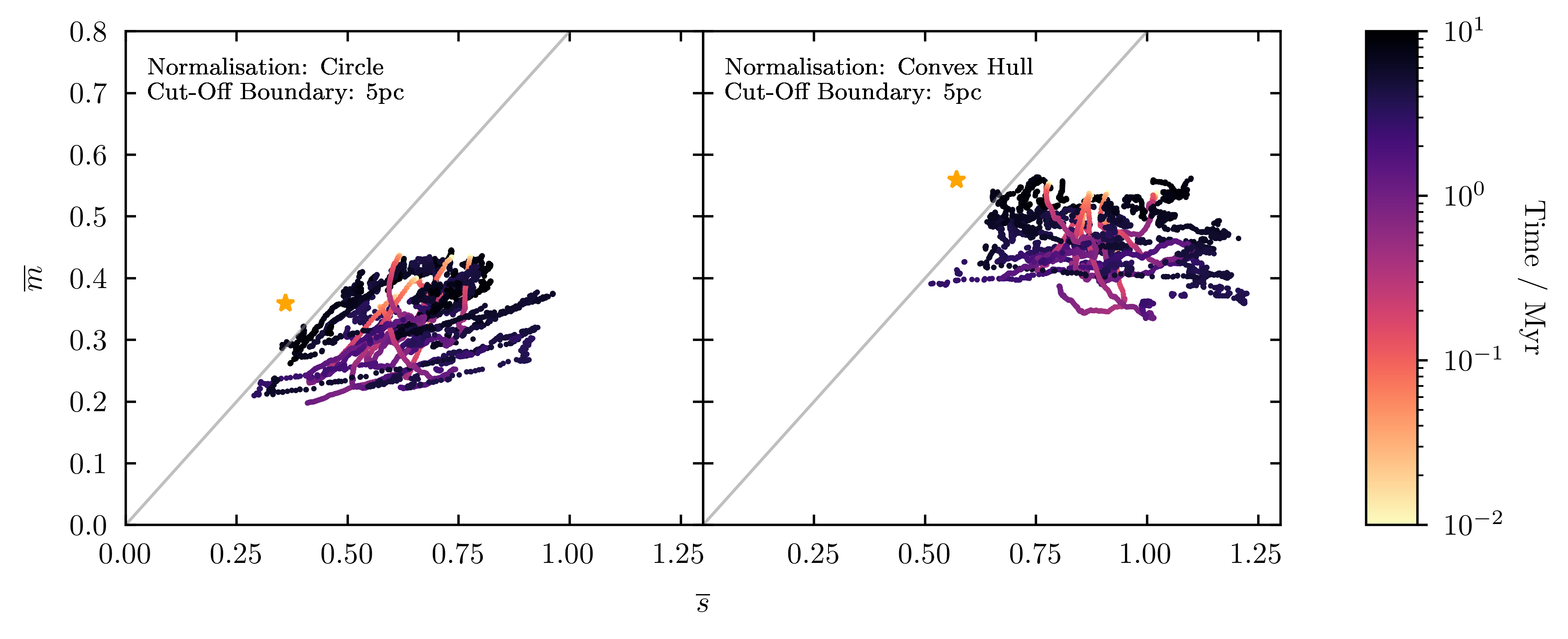}
	\caption{Evolution of $\overline{m}$ and $\overline{s}$ over 10 Myr for an IC~348-like region, with initial $D = 2.0$ and $\alpha = 1.5$.
    Ten realisations of the region are shown for both circular and convex hull normalisations.
    The current values of $\overline{m}$ and $\overline{s}$ for IC~348 (age $\approx3-6$ Myr) are shown as a yellow star.}
    \label{fig:ms_IC348_A15_two}
\end{figure*}

Figure~\ref{fig:ms_IC348_A15_two} shows that the evolutionary tracks are a lot less coherent for the initially supervirial $\alpha = 1.5, D = 2.0$ IC~348-like simulations. There is still an initial drop in $\overline{m}$ during the first $\sim 0.1$ Myr, however there also tends be a significant increase or decrease in $\overline{s}$ during this time which is not seen in the initially subvirial regions. There is therefore no general trend in the realisations' movements across the $\overline{m}-\overline{s}$ plot.

However, these simulations do not significantly cross the $Q = 0.8$ line. This means that the supervirial simulations do not become smooth and centrally concentrated. Instead, they remain substructured throughout the 10 Myr simulated here. This is because these regions immediately expand and therefore there is not enough dynamical mixing to erase the substructure.

Since IC~348 lies above the $Q = 0.8$ line, the $\overline{m}-\overline{s}$ plot would suggest that it is not possible for it to have had supervirial initial conditions.

\subsubsection{Constraints placed on IC~348}
All initial conditions with $\alpha = 0.3$ where a cut-off boundary was imposed are consistent with the current observed values of $\overline{m}$ and $\overline{s}$ for IC~348, between $\sim1-10$ Myr. 

In terms of initial substructure, both our smooth and substructured initial conditions are consistent with observations of IC~348. However, there is evidence that star-forming regions tend to be initially substructured \citep{2004CartwrightWhitworth, 2009SanchezAlfaro, 2010AndreEtAl, 2014AndreEtAl, 2014KuhnEtAl, 2015JaehnigEtAl, 2019ArzoumanianEtAl}. These results would therefore suggest that IC~348 likely had substructured and subvirial initial conditions.

\subsection{NGC~1333} 
\begin{figure*}
	\includegraphics[width=\textwidth]{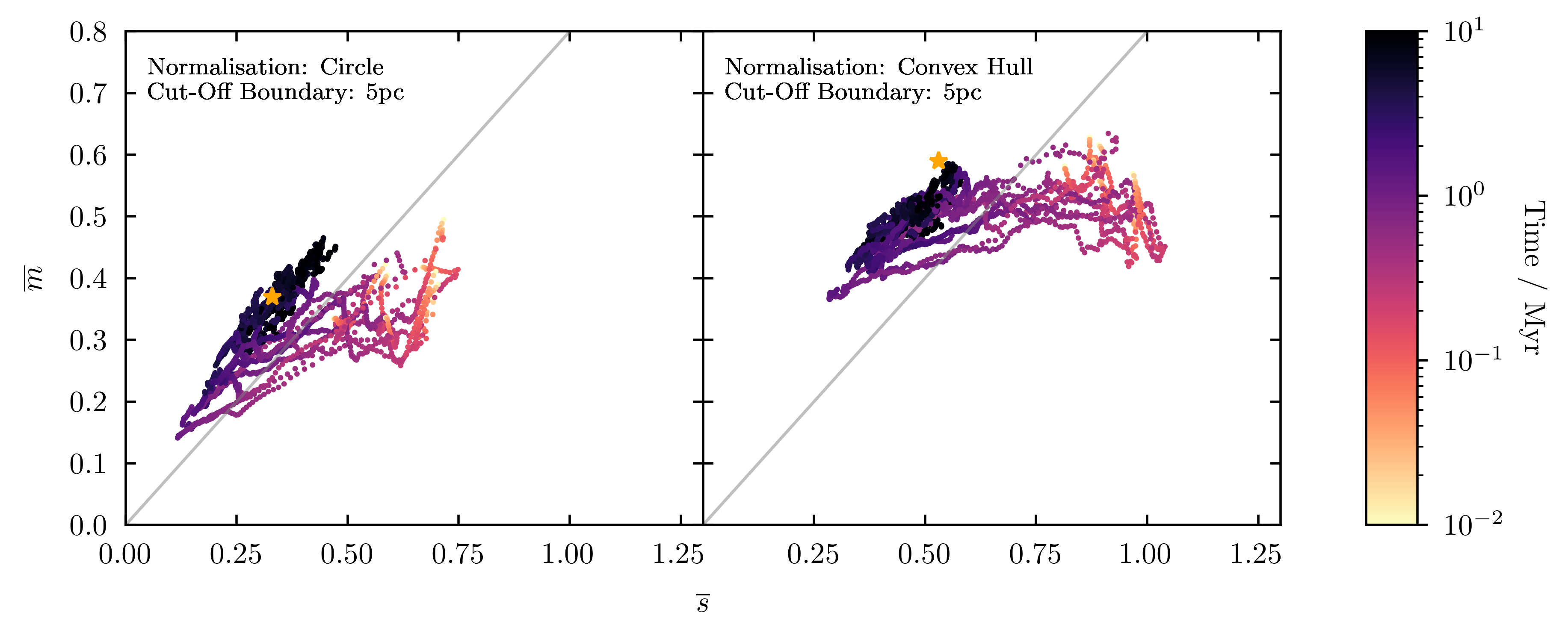}
	\caption{Evolution of $\overline{m}$ and $\overline{s}$ over 10 Myr for an NGC~1333-like region, with initial $D = 2.0$ and $\alpha = 0.3$.
	Ten realisations of the region are shown for both circular and convex hull normalisations.
    The current values of $\overline{m}$ and $\overline{s}$ for NGC~1333 (age $\sim1$ Myr) are shown as a yellow star.
    The evolutionary path is similar to the corresponding IC~348 simulations.}
    \label{fig:ms_NGC1333_D2_two}
\end{figure*}

The evolution of each set of NGC~1333-like simulations follows the same general migration as their corresponding IC~348-like regions, including an evolution across the $\overline{m}-\overline{s}$ plot gap for those that begin substructured. The evolution of the NGC~1333-like region with initial $D = 2.0$ and $\alpha = 0.3$ can be seen in Figure~\ref{fig:ms_NGC1333_D2_two}, which shows the initial drop in $\overline{m}$, followed by a turn-over at $\sim1$ Myr in the same region of the plot as for the IC~348-like regions.

\subsubsection{Constraints placed on NGC~1333}
Like IC~348, NGC~1333 is observed to lie above the $Q = 0.8$ line. Our simulations are therefore inconsistent with NGC~1333 having been initially highly supervirial ($\alpha = 1.5$) if it formed substructured \citep{2004CartwrightWhitworth, 2009SanchezAlfaro, 2010AndreEtAl, 2014AndreEtAl, 2014KuhnEtAl, 2015JaehnigEtAl, 2019ArzoumanianEtAl}.

Figure~\ref{fig:ms_NGC1333_D2_two} shows some of the realisations that are substructured and subvirial to be consistent with the current observed values of $\overline{m}$ and $\overline{s}$ for NGC~1333 at ages greater than $\sim 1$ Myr - slightly older than NGC~1333 is thought to be.

\subsection{ONC} 
\begin{figure*}
	\includegraphics[width=\textwidth]{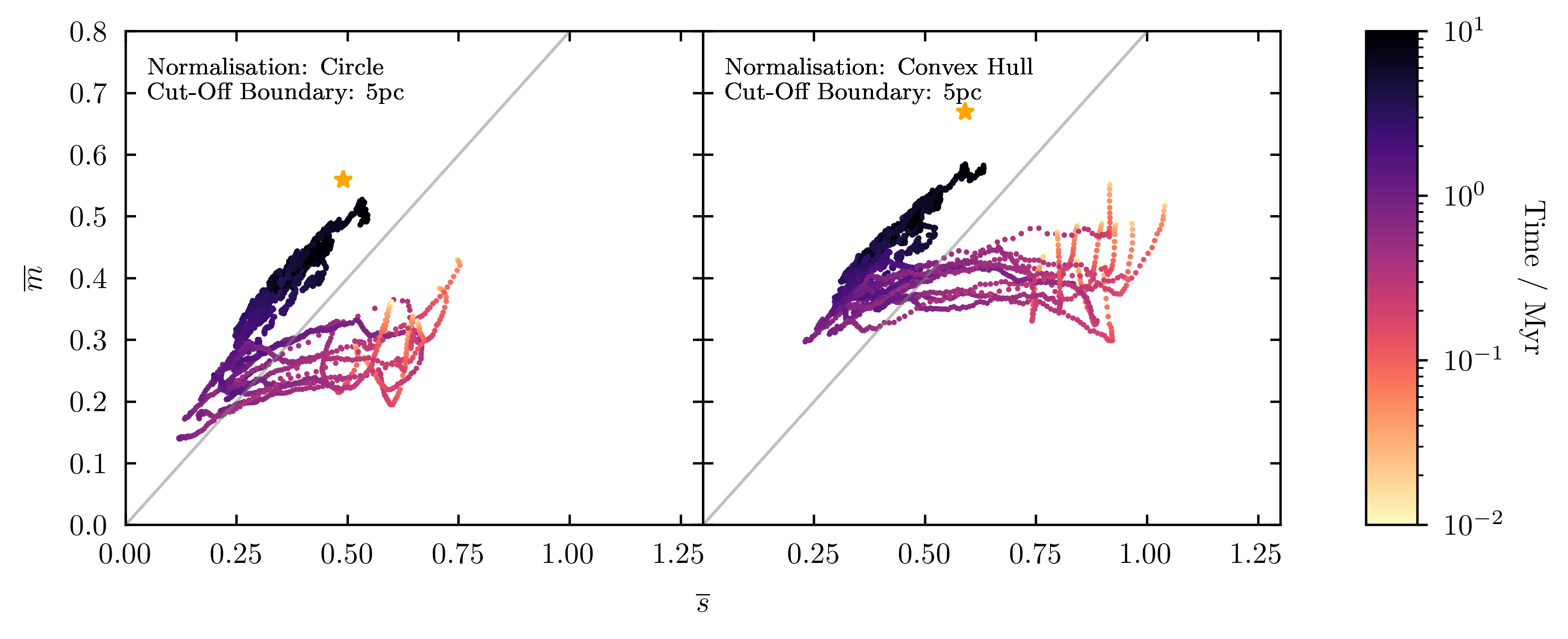}
	\caption{Evolution of $\overline{m}$ and $\overline{s}$ for an ONC-like region, with initial $D = 2.0$ and $\alpha = 0.3$.
	Ten realisations of the region are shown for both circular and convex hull normalisations.
	The current values of $\overline{m}$ and $\overline{s}$ for the ONC (age $\approx1-4$ Myr) are shown as a yellow star.
    The evolutionary path is similar to the corresponding IC~348 and NGC~1333 simulations.}
    \label{fig:ms_ONC_D2_two}
\end{figure*}

The evolution of our ONC-like simulations also follow the same characteristic evolution as each of their corresponding IC~348 and NGC~1333-like regions. This can be seen for the $D = 2.0$ $\alpha = 0.3$ ONC-like simulations in Figure~\ref{fig:ms_ONC_D2_two}.

\subsubsection{Constraints placed on the ONC}
None of our sets of simulations are in agreement with current observations of the ONC, as no realisations reach a high enough value of $\overline{m}$ at any point in their evolution to be consistent with observations. 

This is likely in part due to the lane of dust across the centre of the ONC \citep[shown in figure 3 of][]{1997Hillenbrand}, as the $\mathcal{Q}$-parameter can be affected by significant amounts of extinction \citep{2012ParkerMeyer}.
This dust lane likely excludes a significant number of the central stars and gives a false effect of more substructure.

\begin{figure}
\hspace*{-1.2cm}
		\includegraphics[scale=0.4]{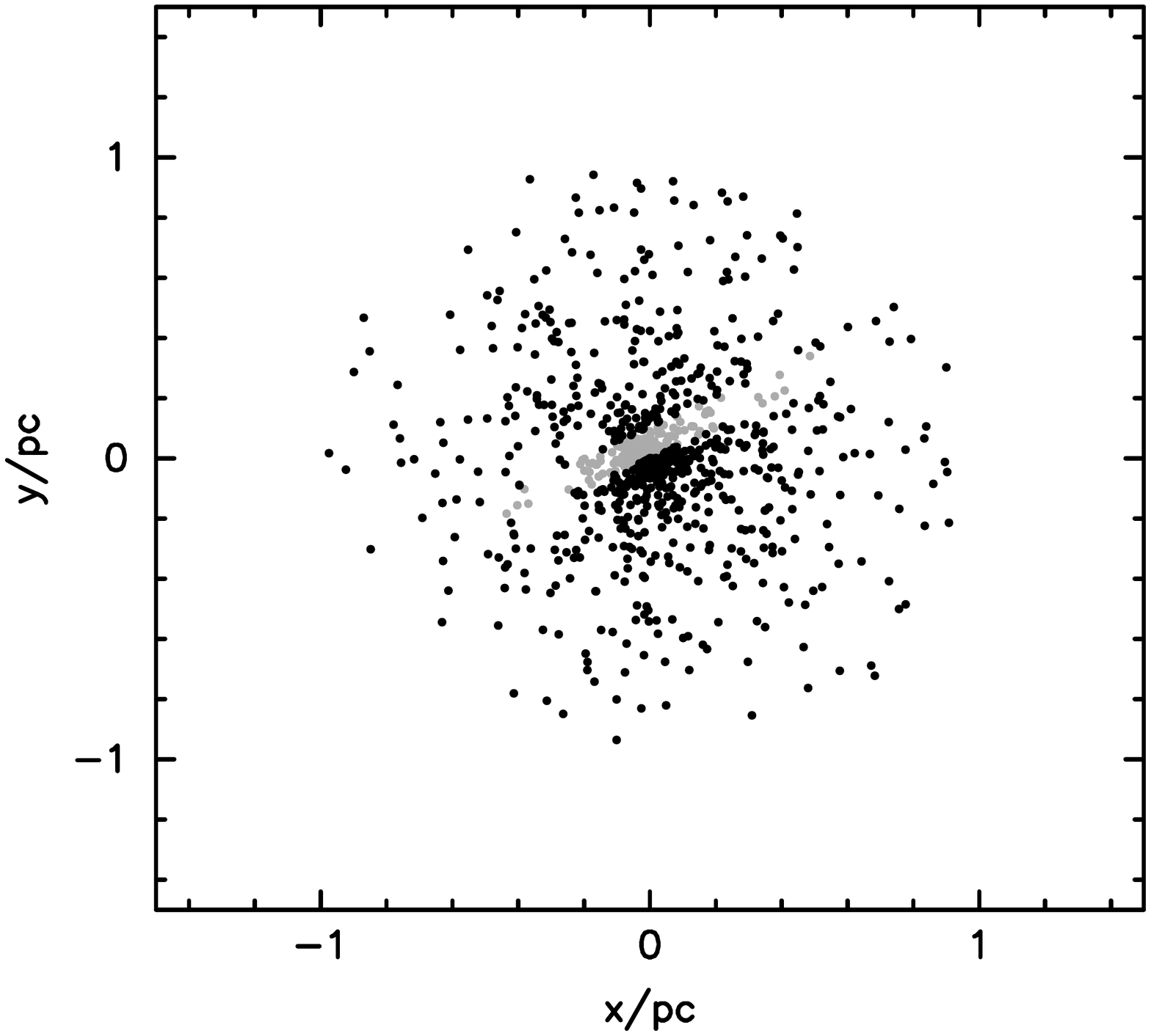}
	\caption{The spatial distribution of a smooth centrally concentrated ONC-like region. The band of excluded stars are shown in light grey.}
    \label{fig:Richards_plot}
\end{figure}

We have tested this by excluding a band of stars for a smooth and centrally concentrated ONC-like region, as shown in Figure~\ref{fig:Richards_plot}. This tends to increase $\overline{m}$ and $\overline{s}$ by $\approx0.04$ and $\approx0.08$ respectively, moving the simulations closer to the observed values for the ONC.

Finally, we note that the ONC is one component in a much larger star-forming region, and it is often unclear where the edge of this cluster lies in relation to other stars in the Orion region. Determining the $\mathcal{Q}$-parameter in star-forming regions with overlapping populations is notoriously problematic, and is worth bearing in mind when interpreting our results. It is also possible that the dynamical evolution of the ONC has been influenced by its surrounding environment (i.e\,\,the gravitational potential of the Orion cloud), something not included in these simulations.

\section{Discussion} 

\subsection{Evolutionary Trends} 
All subvirial regions simulated here evolved from their initial conditions towards the area of the $\overline{m}-\overline{s}$ plot that corresponds to a smooth centrally concentrated region. This transition takes $\sim1$ Myr, which is in agreement with observations of the timescales within which dynamical interactions tend to erase substructure in young star-forming regions \citep{2015JaehnigEtAl}. When a cut-off boundary is used in the analysis, after reaching this area of the plot at $\sim1$ Myr, each realisation then evolves up along the $Q = 0.8$ line - becoming less centrally concentrated for the rest of the 10 Myr simulation.

\subsubsection{Effect of Initial Substructure}
For regions that begin substructured, dynamical evolution takes them across the $\overline{m}-\overline{s}$ plot gap as they transition from fractal to smooth and centrally concentrated. However, for regions which begin smooth, with a fractal dimension of $D = 3.0$, this movement takes them downwards along the $Q = 0.8$ line, meaning that they do not cross the $\overline{m}-\overline{s}$ plot gap. This suggests that, since only regions that begin substructured evolve to populate the $\overline{m}-\overline{s}$ plot gap, star-forming regions which are currently observed to lie in the gap would have been substructured in their past - ruling out smooth initial conditions for regions such as Cha~I and Taurus (Figure~\ref{fig:msplotgap}).

\subsubsection{Effect of Initial virial Ratio}
Our substructured simulations evolve to cross the $\overline{m}-\overline{s}$ plot gap for virial ratios of 0.3 and 1.5. This implies that, as long as a star-forming region begins substructured, it will evolve into the $\overline{m}-\overline{s}$ plot gap regardless of whether it is initially collapsing or expanding.

However, the virial ratio does affect whether an initially substructured cluster will evolve to become smooth and cross the $Q = 0.8$ line, as our supervirial ($\alpha = 1.5$) simulations remain substructured throughout the 10 Myrs of evolution. Since star-forming regions are likely initially substructured \citep{2004CartwrightWhitworth, 2009SanchezAlfaro, 2010AndreEtAl, 2014AndreEtAl, 2014KuhnEtAl, 2015JaehnigEtAl, 2019ArzoumanianEtAl}, this would suggest that supervirial initial conditions can be ruled out for any star-forming region that is observed to be smooth.

\subsection{Problems with the Q-parameter method} 
\label{sec:problems}
Our results show that, although idealised geometries do not populate all of the same areas of the $\overline{m}-\overline{s}$ plot as observed regions, this should not be seen as a problem with either using these as initial conditions or the $\overline{m}-\overline{s}$ plot as a method of analysis, because simulated regions move into the $\overline{m}-\overline{s}$ plot gap as they dynamically evolve. 

However, there are drawbacks to be considered when using the $\mathcal{Q}$-parameter method to analyse dynamical evolution. Some of these affect the interpretation of the results presented here, especially when using an evolutionary track of an $\overline{m}-\overline{s}$ plot to infer the likely past or future evolution of an observed region. 

Figure~\ref{fig:ms_IC348_A15_two} shows that the path which initially supervirial ($\alpha = 1.5$) regions take across the $\overline{m}-\overline{s}$ plot is significantly less predictable than for initially subvirial ($\alpha = 0.3$) regions. This means that this method may be less able to reliably predict likely future or past evolution of observed regions that were initially supervirial. However, as discussed, it is still able to rule out some initial conditions.

Incomplete or inaccurate observational data can also have a significant effect on the ability to compare simulations analysed using this method to observations.
For example, the ages of observed star-forming regions are likely only accurate to a factor of $\sim2$ for those under 10 Myr \citep{2014SoderblomEtAl}.
Taking Figure~\ref{fig:ms_NGC1333_D2_two} as an example, this, combined with the inherent lack of error bars in the $\mathcal{Q}$-parameter method, makes it hard to fully assess whether our simulations are consistent with the observed values of $\overline{m}$ and $\overline{s}$ for NGC~1333.

Values of $\overline{m}$ and $\overline{s}$, and the $\mathcal{Q}$-parameter overall, can also be sensitive to which stars are included or excluded from the analysis. This includes foreground/background stars, extinction and crowding \citep{2012ParkerMeyer}, as well as the chosen cluster radius. This is shown here in Figure~\ref{fig:ms_IC348_D2_all} where, at later times, the evolutionary tracks are significantly changed once a cut-off boundary is used. 

\section{Conclusions}
We have used N-body simulations of young star-forming regions to investigate their dynamical evolution in $\overline{m}$, $\overline{s}$, and the $\mathcal{Q}$-parameter.

Our main results are summarised as follows:
\begin{enumerate}
    \item All of our initially substructured star-forming regions move into the $\overline{m}-\overline{s}$ plot gap. This happens as they dynamically evolve towards more smooth and centrally concentrated distributions over the first $\sim1$ Myr.
    \item This suggests that any star-forming region which is observed to lie in the $\overline{m}-\overline{s}$ plot gap must have been initially substructured, regardless of whether it was initially super or subvirial.
    \item Our initially supervirial substructured simulations do not cross the $Q = 0.8$ line, and therefore never become smooth and centrally concentrated. Since star-forming regions are observed to be initially substructured, this would suggest that a region which is observed to lie above the $Q = 0.8$ line must have been initially subvirial.
    \item All of our subvirial IC~348-like simulations are consistent with IC~348's observed values of $\overline{m}$ and $\overline{s}$, although occasionally at slightly older ages than IC~348 is estimated to be. Our set of simulations suggest that supervirial initial conditions can be ruled out for IC~348. And, since star forming-regions are likely initially substructured, this would therefore imply that IC~348 was substructured and subvirial in its past.
    \item Our NGC~1333-like simulations would also suggest substructured and subvirial initial conditions. However the evolutionary tracks of each realisation that matches NGC~1333 tend to do so at an age of several Myr, compared to NGC~1333's estimated age of $\sim1$ Myr.
    \item None of our simulations were consistent with observations of the ONC, as none of them populate the same area of the $\overline{m}-\overline{s}$ plot. This is likely, at least in part, due to the ridge of extinction across the middle of the ONC which has the effect of causing regions to appear more substructured.
\end{enumerate}

As with any analysis technique, there are drawbacks with using the $\mathcal{Q}$-parameter. 
However, our simulations show that box fractal regions will dynamically evolve into the $\overline{m}-\overline{s}$ plot gap - populating the area where some observed star-forming regions lie (e.g. Cha~I and Taurus, as shown in Figure~\ref{fig:msplotgap}). 
Star-forming regions can therefore be consistent with having evolved from these fractal geometries in terms of their values of $\overline{m}$ and $\overline{s}$. 

Our results show that observed values of $\mathcal{Q}$,  $\overline{m}$ and $\overline{s}$ should not be directly compared to idealised geometries, as these synthetic box fractals and smooth, centrally concentrated profiles have not undergone any dynamical evolution, nor are they subjected to the observational biases of real star-forming regions. Instead, $\mathcal{Q}$,  $\overline{m}$ and $\overline{s}$ should only be used to determine the degree to which a star-forming region is either  spatially substructured or smooth and centrally concentrated. 

\section*{Acknowledgements}

We thank the anonymous referee for their helpful comments and suggestions. ECD acknowledges support from the UK Science and Technology Facilities Council in the form of a PhD studentship.
RJP acknowledges support from the Royal Society in the form of a Dorothy Hodgkin Fellowship.


\bibliographystyle{mnras}
\bibliography{references}




\bsp	
\label{lastpage}
\end{document}